\documentclass[10pt]{article}
\usepackage[centertags]{amsmath}
\usepackage{graphicx}
\usepackage{graphics}
\newcommand{\be}{\begin{equation}}
\newcommand{\ee}{\end{equation}}
\usepackage[centertags]{amsmath}
\usepackage{amsfonts}

\normalbaselines

\begin{document}
\title{\bf Non-Abelian Vortices  on the Torus}
\author{\normalsize G.S.~Lozano$^a$\thanks{Associated with CONICET}\,,
D.~Marqu\'es$^{b, c \, *}$ and F.A.~Schaposnik$^{b,
c}$\thanks{Associated with CICBA}
\\
{\small \it $^a$Departamento de F\'\i sica, FCEyN, Universidad de
Buenos Aires, 1428, Buenos Aires, Argentina}
\\
{\small\it $^b$Departamento de F\'\i sica, Universidad Nacional de
La Plata, C.C. 67, 1900 La Plata, Argentina}
\\
{\small\it $^c$CEFIMAS-SCA, C1059ABF, Buenos Aires, Argentina} }

\maketitle
\begin{abstract}
We study   periodic arrays of non-Abelian vortices in an
$SU(N)\times U(1)$ gauge theory  with  $N_f$ flavors of
fundamental matter multiplets. We carefully discuss the
corresponding twisted boundary conditions on the torus and propose
an ansatz to solve the first order Bogomolnyi equations which we
find by looking to   a bound
 of the energy.  We solve the equations numerically and construct explicit vortex solutions.
\end{abstract}

\section{Introduction}

Vortices -string-like configurations with quantized magnetic flux-
 have been introduced in the context of High Energy Physics more
than 30 years ago in the pioneering work of Nielsen and Olesen
\cite{N} on the Abelian Higgs model.  Very soon it was shown that
in this model the equations of motion can be reduced to first
order Bogomolnyi-Prasad-Sommerfield (BPS) or self-dual equations
\cite{bogo},\cite{deVega}, explicit numerical solutions were
obtained and the connection with supersymmetry was signaled
\cite{deVega}.

The formal proof of existence of a general $n$-vortex
configurations for the BPS equations in $R^2$ ($n$ being the
vorticity or number of flux-quanta) was later given
in~\cite{taubes}. Although the existence of vortex solutions for
the BPS equations on other (Kahler) manifolds was shown in
\cite{bradlow}, it was only very recently that Gonzalez Arroyo and
Ramos \cite{GaR} presented a simple method to construct explicit
numerical solutions on the bi-dimensional torus ${\cal{T}}^{2}$,
the simplest compact manifold. The case of the torus is
particularly interesting as it corresponds to the study of
periodic field configurations, leading in this case to
vortex-lattice arrays (a situation that most often arise in
condensed matter problems). The torus also provides the most
natural (long-distance) regularization of $R^2$, so even for the
cases in which the problem of interest is set in the plane, it is
important to have a well controlled method that allows us to study
the asymptotic infinite area case. For instance, the torus
provides a natural set up to study numerically
non-cylindrically-symmetric multi vortex configurations (the case
of $n$ superimposed vortices or ``giant" vortex  can be more
easily treated on a disc).

 Self-dual equations are of course simpler to study than
the Euler-Lagrange equations but, more importantly, it is their
connection to supersymmetry and their relevance to the understanding
of non-perturbative phenomena in field theories and string theory
that have triggered so many investigation in the last years (for
recent reviews on the subject see \cite{lecTong}, \cite{fas}) .

It was   shown  long ago that non-Abelian gauge models  also have
vortex solutions \cite{dVS2} (see also references in
\cite{lecTong}-\cite{fas}). More recently much attention has been
devoted to the analysis of certain ${\cal N}=2$ supersymmetric
theories where non-Abelian vortices have been presented and used to
construct low-energy effective actions related to string dynamics
\cite{Hanany1}-\cite{mod2}. With this in mind, local and semi-local
vortex solutions have been studied in models in which Yang-Mills
fields are coupled to matter fields with different numbers of colors
and flavors (see \cite{revshif} for a review).

In this work we  construct non-Abelian vortices in a gauge theory
defined on the torus. Motivated by the great interest that the type
of theories considered in  \cite{Hanany1}-\cite{mod2} has received
recently, we  consider a model with gauge group $SU(N)\times U(1)$
with $N_f$ flavors of fundamental matter multiplets. After defining
the model in section 2, we  discuss in section 3 the appropriate
boundary conditions on the torus which will lead to the non-trivial
magnetic flux associated to the vortices. In section 4 we obtain a
bound for the energy per unit length of the static vortices (the
string tension) and from it we derive the first order Bogomolnyi
equations (which correspond to a minimum of the energy). We present
vortex solutions to these equations  in section 5 for the case
$N=N_f=2$ and then extend the analysis to general $N$ in section 6.
We summarize and discuss our results in section 7.

\section{The model}
We consider the bosonic sector of an ${\cal N} = 2$ supersymmetric
Yang-Mills-Higgs with $SU(N)\times U(1)$ gauge group and $N_f$
flavors of fundamental matter multiplets, described by the action
\cite{SY}

\begin{eqnarray}
{\cal S} = \int d^4x &&\!\!\!\!\!\!\!\!\!\!\!\!
 \left[\frac{1}{4g^2}F_{\mu\nu}^m
F^{m\,\mu\nu}  + \frac{1}{4e^2} F_{\mu\nu}^0 F^{0\,\mu\nu}  +
\left(D_\mu\phi^f\right)^\dag\left(D^\mu\phi^f\right) \right.\nonumber\\
&&\!\!\!\!\!\!\!\!\left. - \frac{g^2}{2}\mid\phi^\dag_f t^m\phi^f\mid^2 -
\frac{e^2}{4N}\left(\phi^\dag_f\phi^f - N\xi\right)^2 \right]\
.\label{accion}
\end{eqnarray}
In the present case, coordinates $x_1, x_2$ are defined on the
two-torus ${\cal T}^2$ of size $L_1\times L_2$. The scalar matter
fields are denoted by $\phi^f$ with $f$ the flavor index, $f =
1,\dots, N_f$; $\mu, \nu = 0, 1, 2, 3$ are Lorentz indices and $m =
1, ..., N^2-1$ is an internal space index of $SU(N)$. We indicate
with a superindex $0$ the $U(1)$ components of the gauge fields. The
$SU(N)$ generators $t^m$ are taken to be anti-hermitian with the
following normalization
\begin{equation}
Tr \left(t^m t^n\right)=-\frac{1}{2}\delta^{mn}\ ,\ \ \ \
\left(t^m\right)^i_j \left(t^m\right)^k_l=-\frac{1}{2}\delta^i_l
\delta^k_j + \frac{1}{2N} \,\delta^i_j \delta^k_l\ , \ \ \ \ [t^m,
t^n] = f^{mnl} t^l\ .
\end{equation}
Here $i,j = 1,2,\ldots,N$ are indices in the fundamental
representation of $SU(N)$.
The covariant derivative reads
\begin{equation}
D_\mu\phi^f = \partial_\mu \phi^f - A_\mu^m t^m \phi^f -
 A_\mu^0 t^0\phi^f
\end{equation}
and we    take  the $U(1)$ generator $t^0 = (i/\sqrt{2N}) I $.
Field strengths associated with gauge fields $A_\mu = A_\mu^m t^m
$ and $A_\mu^0$ are defined as
\begin{eqnarray}
F_{\mu\nu}  &=& \partial_\mu A_\nu  - \partial_\nu A_\mu  - [A_\mu,A_\nu] =
\frac{i}{\sqrt{2N}} F^0_{\mu\nu} I + F^m_{\mu\nu}t^m \nonumber\\
F^0_{\mu\nu}  &=& \partial_\mu A^0_\nu  - \partial_\nu
A^0_\mu\nonumber\\
F^m_{\mu\nu}  &=& \partial_\mu A^m_\nu  - \partial_\nu A^m_\mu -
f^{mnl}A_\mu^n A_\nu^l\ .
\end{eqnarray}

The two last terms in (\ref{accion}) are  responsible for gauge
symmetry braking. The last one, containing the Fayet-Iliopoulos
parameter $\xi$, forces $\phi^f$ to develop a vacuum expectation
value, while the last but one forces the VEV to be diagonal. We
shall be interested in the case $N_f \geq N$ since (when $N_f<N$
there is spontaneous supersymmetry breaking \cite{lecTong}).

Up to gauge transformations, the minimum of the potential, which we
call $\Phi_0$, can be written as an $N\times N_f$ matrix,
\begin{equation}
\Phi_0 = \sqrt{\xi}\left(\begin{matrix}1 &  & 0 & 0 & \dots & 0\\
 & \ddots & & \vdots & \ddots & \vdots \\
 0 & & 1& 0 & \dots & 0\end{matrix}\right)\ .
\end{equation}
Here rows correspond to the $N$ colors and columns to the $N_f$
flavors.  Such a minimum breaks the symmetry $U(1)\times SU(N)\times
SU(N_f)$ down to a global remanent symmetry $SU(N)_{C+F}$,
\begin{equation}
U(1)\times SU(N)\times SU(N_f) \ \ \ \to \ \ \ SU(N)_{C + F}\ .
\end{equation}
 The global $SU(N)_{C+F}$ transformations leaving the vacuum invariant act on $\Phi_0$ according to
\begin{equation}
\Phi_0 = U^{-1} \Phi_0 \left(\begin{matrix} U \\ 0
\end{matrix}\right)\ ,
\end{equation}
where $U$ is a global $SU(N)$ matrix, and $0$ is the null $(N_f - N)
\times N$ matrix.

\section{Boundary conditions}

Since we are working in the two-torus $(x_1,x_2)\subset {\cal T}^2$,
the
  gauge fields $A_\mu$ and the  matter multiplets $\phi^f$
 must obey
periodic boundary conditions on ${\cal T}^2$ modulo  gauge
transformations
\begin{eqnarray}
A_\mu(x_i + L_i) &=& U_i A_\mu U_i^{-1} + \partial_\mu U_i
U_i^{-1}\label{gaugetran}\\
\phi^f(x_i + L_i) &=& U_i \phi^f \ ,\label{phitran}
\end{eqnarray}
where $U_i  \in  U(1)\times SU(N)$ ($i=1,2$)  are the transition
functions. Consistency of  equation (\ref{gaugetran}) for the gauge fields leads to the
following relation for the transition functions
\begin{equation}
U_2(x_1 + L_1, x_2) U_1(x_1, x_2) = U_1(x_1, x_2 + L_2) U_2(x_1,
x_2)\ \Omega\ .\label{consistency}
\end{equation}
Here $\Omega$ is  an element of the $Z_N\times U(1)$ center of
$U(N)$ \cite{GA2} which labels inequivalent topological sectors.
Its presence is due to the fact that transition functions $U_i$ can
be defined in (\ref{gaugetran}) modulo an element of the $U(N)$
center. However, in the present model there are   matter fields
in the fundamental representation
and one has also to check consistency of equation (\ref{phitran}).
In this case one finds
\begin{equation}
U_2(x_1 + L_1, x_2) U_1(x_1, x_2) = U_1(x_1, x_2 + L_2) U_2(x_1,
x_2)\ ,\label{consistencyy}
\end{equation}
so that $\Omega$ in (\ref{consistency}) should be taken as the unity
matrix, $\Omega = I$.

In order to find the solution to equation (\ref{consistencyy}) it
will be convenient to construct the following linear combination of
elements in the Cartan subalgebra of $U(N)$
\begin{eqnarray}
\tau_0 &=& \frac{i}{\sqrt{2N}}\ {\rm diag}(1,\dots, 1) \nonumber\\
\tau_1 &=& \frac{i}{\sqrt{2N(N-1)}}\ {\rm diag}(- (N-1),1, \dots, 1)
\nonumber\\
&\vdots& \nonumber\\
\tau_N &=& \frac{i}{\sqrt{2N(N-1)}}\ {\rm diag}(1, \dots, 1, -
(N-1))\ .\label{cartangen}
\end{eqnarray}
We then define a $q$-elementary
transition function solution $U_i^{(q)}(x_1,x_2)$ as that which is
generated by $\tau_0$ (which is proportional to the identity)  and
$\tau_q$,
\begin{eqnarray}
U_1^{(q)}(x_1, x_2) &=& \prod_m\exp \left(-  \frac{\tau_m \gamma_m}{2}
\frac{  x_2}{L_2}\right) \;\;\; \nonumber\\
U_2^{(q)}(x_1, x_2) &=& \prod_m\exp \left(\frac{\tau_m \gamma_m}{2}
\frac{  x_1}{L_1}\right)\ , \label{transition}
\end{eqnarray}
with
\begin{eqnarray}
\gamma_0 &=& - 2\pi \sqrt{\frac{2}{N}}  \nonumber\\
\gamma_q &=& 2\pi \sqrt{\frac{2(N-1)}{N}}\nonumber\\
\gamma_i &=& 0 \;\;\; \forall \; i \ne 0,q\ . \label{gamas}
\end{eqnarray}
Using this transition function, we shall be able to construct
elementary vortex solutions in the torus which in the $L_1,L_2 \to
\infty$ limit  reduce to the elementary vortex solutions in
\cite{Auzzi},\cite{GSY}. Such elementary vortices have a quantum of
magnetic flux.

Vortices with higher units of magnetic flux are constructed by
considering transition functions which are products of elementary
transition functions. These general transition functions are
connected with those introduced by 't Hooft
\cite{GA2},\cite{tHooft}. In fact, a general transition
function with $n_0$ units of magnetic flux reads
\begin{eqnarray}
\gamma_0 &=& - 2\pi \sqrt{\frac{2}{N}}\ n_0  \nonumber\\
\gamma_q &=& 2\pi \sqrt{\frac{2(N-1)}{N}}\ n_q\ , \ \ \ \ \ \ q =
1,\dots,N\ ,\label{quince}
\end{eqnarray}
where
\begin{eqnarray}
n_0 = \sum_{q = 1}^N \ n_q\ .\label{enecero}
\end{eqnarray}

With the transition functions that we have defined, conditions
(\ref{gaugetran}) read
\begin{eqnarray}
A_1(x_1 + L_1, x_2) &=& A_1(x_1, x_2)\nonumber\\
A_1(x_1, x_2 + L_2) &=& A_1(x_1, x_2) -
\frac{1}{2L_1}\sum_m\gamma_m\tau_m\nonumber\\
A_2(x_1 + L_1, x_2) &=& A_2(x_1, x_2) +
\frac{1}{2L_2}\sum_m\gamma_m\tau_m\nonumber\\
A_2(x_1, x_2 + L_2) &=& A_2(x_1, x_2) \ ,
\end{eqnarray}
and  can be written as a sum of a periodic function
$\tilde{A_i}$ in ${\cal T}^2$ plus a known function
\begin{equation}
A_i(x_1, x_2) = \tilde{A}_i(x_1, x_2) +
\frac{1}{2L_1L_2}\sum_m\gamma_m\tau_m \epsilon_{ij} x_j\ ,
\label{16}
\end{equation}
so that the field strength takes the form
\begin{eqnarray}
F_{ij} &=& \tilde{F}_{ij} -
\frac{\epsilon_{ij}}{L_1L_2}\sum_m\gamma_m\tau_m
\nonumber\\
\tilde{F}_{ij} &=& \partial_i\tilde{A}_j - \partial_j \tilde{A}_i -
\left[\tilde{A}_i, \tilde{A}_j\right]\ .
\label{efes}
\end{eqnarray}
The magnetic flux can be defined from (\ref{efes}) as
\begin{equation}
\Phi_{mag} = -i{\rm Tr} \int_{{\cal T}^2} d{\cal T} F_{12} = -
\sqrt{\frac N2} \gamma^0 = 2\pi\ n_0  .\label{18}
\end{equation}

In analogy with what we did for the gauge field, we shall propose
an ansatz for matter fields factoring out in each multiplet a
particular $N\times N$ matrix $\chi = \left(\chi_{ij}(x_1,
x_2)\right)$ satisfying the twisted boundary conditions, times a
scalar multiplet carrying both color and flavor indices,
$\Lambda_i^f(x_1, x_2)$ obeying periodic boundary conditions
\begin{equation}
\phi^f(x_1, x_2) =  \chi(x_1, x_2)  \Lambda^f (x_1, x_2)\ .
\label{Lambda}
\end{equation}
Given conditions (\ref{phitran}),   $\chi(x_1, x_2)$ should satisfy
\begin{eqnarray}
\chi(x_1 + L_1, x_2) &=& \prod_m\exp\left(-\frac{\tau_m\gamma_m}{2}
\frac{x_2}{L_2} \right)\chi(x_1, x_2)\nonumber\\
\chi(x_1, x_2 + L_2) &=&
\prod_m\exp\left(\frac{\tau_m\gamma_m}{2}\frac{x_1}{L_1} \right)\chi(x_1,
x_2)\ .\label{bchiggs}
\end{eqnarray}
A solution to these equations is given by
\begin{equation}
\chi(x_1, x_2)=\prod_{m= 0}^N \exp\left(-\frac{(x_1 + i x_2)
x_2}{{L_1 L_2} }
\frac{\gamma_{m}\tau_{m}}{2}\right)\Theta^{m}\left(x_1,
x_2\right)\label{chidef}
\end{equation}
with
\begin{equation}
\Theta^{m}(x_1, x_2) = \prod_{n = 1}^{\mid
n_{m}\mid}\theta_3\left(i\ \frac{\gamma_{m}\tau_{m}}{2 n_{m}}
\frac{(x_1 + i x_2 + a_n^{m})}{L_1}\ \mid
-\frac{\gamma_{m}\tau_{m}}{2\pi n_{m}}\frac{L_2}{L_1} \right)\ .
\end{equation}
Here, $n_{m}$ is the charge associated to the $m$-th direction
of the Cartan subalgebra as defined in (\ref{cartangen}), and
\begin{equation}
\theta_3(z\mid \tau) = \sum_l e^{i\pi \tau l^2 + 2i l z}\ ,
\label{theta3}
\end{equation}
is the Riemann Theta function. The complex coefficients
$a_n^{m}$ satisfy the conditions
\begin{equation}
\sum_{n = 1}^{n_{m}} a_n^{m} = 0\ ,\label{constraint}
\end{equation}
and determine the position of the vortices.

In the next sections, $\Lambda^f(x_1, x_2)$ in (\ref{Lambda})
together with $\tilde A_i(x_1, x_2)$ will be determined from the
equations of motion.

\section{Bogomolnyi equations}

We are interested in infinitely long ($x_3-$independent)
 static configurations which
extremize the tension (energy per unit length),
\begin{equation}
T = \int_{{\cal T}^2} d{\cal T} \left[\frac{1}{4g^2}F_{ij}^m
F^{ij}_m + \frac{1}{4e^2} F_{ij}^0 F^{ij}_0 + \mid D_i\phi^f\mid^2
+ \frac{g^2}{2}\mid\phi^\dag_f t^m\phi^f\mid^2 +
\frac{e^2}{4N}\left(\phi^\dag_f\phi^f - N\xi\right)^2 \right]\ ,
\end{equation}
where $d{\cal T} = dx^1 dx^2$ is the integration measure over the
torus ${\cal T}^2$.

Since  action (\ref{accion}) is the  purely bosonic part of an ${\cal
N}=2$ supersymmetric action, coupling constants and the form of
the potential are automatically adjusted \cite{ENS} so that
 Bogomolnyi completion can be performed \cite{bogo}. Indeed, using the
 relation
\begin{equation}
\mid D_i\phi^f\mid^2 = \mid (D_1 \pm i D_2 )\phi^f\mid^2 \mp i\
\frac{1}{2}\ \phi_f^\dag\ F_{ij}\ \phi^f \epsilon_{ij} + {\rm td}
\end{equation}
with     ``td'' a total derivative term, we can write  the energy
per unit length as
\begin{eqnarray}
T = \int_{{\cal T}^2} d{\cal T}\!\!\!\!\!\!\!\!\!\!\!\!&&\left[\left(\frac{1}{2g}F^m_{ij}
\mp i  \frac{g}{2} \phi^\dag_f t^m\phi^f
\epsilon_{ij}\right)^2 + \left(\frac{1}{2e}F^0_{ij} \pm
\frac{e}{\sqrt{8N}} \left(\phi^\dag_f \phi^f -
N\xi\right)\epsilon_{ij}\right)^2
\right.\nonumber\\
&&\left. + \frac{1}{2}\mid \left(D_i\pm
i \epsilon_{ij}D_j\right)\phi^f \mid^2 \pm  \xi
\sqrt{\frac{N}{2}} F^0_{12}\right]\ .\label{energy}
\end{eqnarray}
Using equation (\ref{18}) we find that the tension is bounded by
\begin{equation}
T \geq   2\pi  \mid n_0 \mid \xi \, , \;\;\; n_0    \in Z
\end{equation}
with $n_0$ defined in
(\ref{enecero}). This bound is saturated whenever the following
 Bogomolnyi equations hold
\begin{eqnarray}
F^m_{ij} &=& \pm i  g^2 \phi^\dag_f
t^m\phi^f \epsilon_{ij} \label{Bogequa1}\\
F^0_{ij} &=& \mp \frac{e^2}{\sqrt{2N}}\left(\phi^\dag_f
\phi^f - N\xi\right)\epsilon_{ij}\label{Bogequa2}\\
D_i \phi^f &=& \mp i \epsilon_{ij}D_j\phi^f \
.\label{Bogequa3}
\end{eqnarray}
As we already mentioned,  solutions to these equations will also
satisfy the second order Euler-Lagrange equations of motion. For
definiteness we shall choose the upper sign in these equations (the
other choice can be handled analogously).

We shall look for solutions to the equations
(\ref{Bogequa1})-(\ref{Bogequa3}) subject to the boundary conditions
discussed in the previous section.
We start from eq.(\ref{Bogequa3}),
\begin{equation}
(D_1 + i D_2)\phi^f = 0\ ,\label{Bogeq1}
\end{equation}
and   write $\phi^f(x_1, x_2)$ in the form (\ref{Lambda})
\begin{equation}
\phi^f(x_1, x_2) =  \chi(x_1, x_2)  \Lambda^f (x_1, x_2)\ .
\label{Lambdadenuevo}
\end{equation}
We have already found the explicit form for $\chi$ (see
eq.(\ref{chidef})), which was obtained by fulfilling the required
boundary conditions. We shall now determine $\Lambda^f$ so that
$\phi^f$ in (\ref{Lambdadenuevo}) satisfies Bogomolnyi equation
(\ref{Bogeq1}). To this end it is convenient to write
\begin{equation}
\Lambda^f (x_1, x_2)= M (x_1, x_2)\ {\cal P}^f \ . \label{lambda2}
\end{equation}
Here  ${\cal P}^f$ is a constant multiplet carrying both color and
flavor indices, while $M(x_1, x_2) \in SU(N) \times U(1)$ is a
diagonal hermitian $N \times N$ periodic matrix,
\begin{equation}
M(x_1, x_2) = e^{i \eta_m(x_1, x_2)\tau_m} \label{eme}
\end{equation}
where functions $\eta_m(x_1, x_2)$  are real and periodic and will be
determined through the remaining Bogomolnyi equations, equations
(\ref{Bogequa1}) and (\ref{Bogequa2}). Now, if we write $\tilde{A}_i$ in (\ref{16}) in terms of
 matrix $M$ in the form
\begin{equation}
\tilde{A}_i = i \varepsilon_{ij}\ \partial_j M M^{-1}\
,\label{Atilde}
\end{equation}
 Bogomolnyi equation (\ref{Bogeq1}) imposes to $\chi$  the condition
\be
 \partial_{\bar{z}}\chi + \frac{i}{4L_1L_2}\ z\ \gamma_m\tau_m\ \chi = 0\ ,
\label{z} \ee where we have defined $z = x_1 + i x_2$. Remarkably,
eq.(\ref{z})   is automatically satisfied by $\chi$ as defined in
 eq.(\ref{chidef}).
Hence, the only remaining task in order to have a complete solution
to the Bogomolnyi equations is to determine $M$ and $P_f$.

\section{Elementary $U(1)\times SU(2)$ non-Abelian vortex solutions in $T^2$}

In this section we complete the construction of string like
solutions to the Bogomolnyi equations for the simplest $U(1)\times
SU(2)$ gauge group case, leaving for the next section the extension
to the general $U(1)\times SU(N)$ case. We shall consider the same
number of flavors and colors, $N = N_f = 2$. We shall also restrict
the analysis to the case of an elementary non-Abelian string
solution with $(n_1, n_2) = (1,0)$. The case in which $(n_1, n_2) =
(0,1)$ is completely analogous.

We have seen from the boundary conditions that the Higgs field
 $\phi^f$   can be factorized as a product of two functions,
one ($\chi$) satisfying non-trivial boundary conditions, the other
($\Lambda^f$), a strictly periodic function which remains to be
computed. In the $N=2$
case $\chi$ is a $2\times 2$ matrix satisfying (\ref{bchiggs})
\begin{eqnarray}
\chi(x_1 + L_1, x_2) &=& \left(\begin{matrix}e^{i\pi
\frac{x_2}{L_2}}& 0 \\ 0 & 1
\end{matrix} \right)
\chi(x_1, x_2)\nonumber\\
\chi(x_1, x_2 + L_2) &=& \left(\begin{matrix}e^{-i\pi
\frac{x_1}{L_1}}& 0 \\ 0 & 1
\end{matrix} \right)\chi(x_1,
x_2)\ .
\end{eqnarray}
Then, we can write $\chi$ as
\begin{equation}
\chi(x_i) = \left(\begin{matrix}\chi_{11}(x_1, x_2)& 0\\0 &
\chi_{22}(x_1, x_2)\end{matrix}\right)
\end{equation}
with $\chi_{11}$ satisfying twisted boundary conditions
\begin{eqnarray}
\chi_{11}(x_1 + L_1, x_2) &=& e^{i\pi \frac{x_2}{L_2}} \chi_{11}
(x_1 ,
x_2) \nonumber\\
\chi_{11}(x_1, x_2 + L_1) &=& e^{-i\pi \frac{x_1}{L_1}}
\chi_{11}(x_1 , x_2) \label{nombre}
\end{eqnarray}
and $\chi_{22}$ periodic in ${\cal T}^2$.

These are very similar to the  boundary conditions that arise in
the Abelian Higgs model in the torus and this is the reason why
construction of the solutions will closely follow \cite{GaR}.
Indeed, as in the Abelian case, we can find a solution to
eqs.(\ref{nombre})   in the form
\begin{equation}
\chi_{11}(x_1,x_2) = \exp\left(i\pi \frac{(x_1 + i x_2) x_2}{{L_1
L_2} } \right) \Theta(x_1,x_2)\ ,\label{chi1}
\end{equation}
with
\begin{equation}
\Theta(x_1, x_2) = \theta_3\left(\pi \frac{x_1 + i x_2}{L_1}\ \mid
i\frac{L_2}{L_1}\ \right)\ . \label{thetaa}
\end{equation}
Here $\theta_3$ is the Riemann theta function already defined in
(\ref{theta3}). Then, $\chi$ can be simply written as
\begin{equation}
\chi(x_1, x_2) = \left(\begin{matrix} \exp\left(i\pi \frac{(x_1 +
i x_2) x_2}{{L_1 L_2} }
\right) \Theta(x_1, x_2) & 0 \\
0 & 1
\end{matrix}\right)\ .
\end{equation}
Without loosing generality we choose $\chi_{22} = 1$ and will
accommodate $\Lambda^f$ so that $\phi^f$ fulfills the appropriate
boundary conditions.

Concerning the gauge field, we shall take  ansatz (\ref{Atilde})
choosing $M$ (see eq.\ref{eme})  in the form
\begin{equation}
M(x_1, x_2) = \exp\left( i \eta_0\tau_0 +
i \eta_1 \tau_1  \right)  \label{emeesp}
\end{equation}
With this choice one can see that the only non-zero components
of the gauge field are $A_i^0$ and $A_i^1$.

 As explained at the end of the previous section, the
ansatz{\ae} (\ref{Lambda}) and (\ref{Atilde}) automatically solve
one of the Bogomolnyi equations, namely eq.(\ref{Bogequa3}). The
problem is then reduced to solving the remaining two equations,
eqs.(\ref{Bogequa1})-(\ref{Bogequa2}),
\begin{eqnarray}
F_{ij}^0 &=& - \left(\phi^\dag_f \phi^f -
2\right)\epsilon_{ij}\label{bogeqs1}\\
F^3_{ij} &=& i 2 \kappa^2\phi^\dag_f t^3 \phi^f\epsilon_{ij}\ .
\label{bogeqs11}
\end{eqnarray}
where $\kappa$ is the ratio of the coupling constants, $\kappa =
g/e$. Here we have scaled coordinates and fields in the form
 \be
x_i \to \sqrt{\frac{2}{e^2\xi}}\, x_i \; , \;\;\;  A_i^m \to
\sqrt{\frac{e^2\xi}{2}}\, A_i^m  \; , \;\;\; \phi^f \to \sqrt \xi \,
\phi^f \ .\ee
Note that periods $L_1,L_2$ are also rescaled according to
\be L_i
\to \sqrt{\frac{2}{e^2\xi}}\, L_i \;
\ee

Recalling equation (\ref{lambda2}), we write
\begin{equation}
\phi^f(x_1, x_2) = \chi(x_1, x_2) M(x_1, x_2) {\cal P}^f\ .
\end{equation}
where the ${\cal P}^f$ multiplets can be chosen in the form
\begin{equation}
{\cal P}^1 = \left(\begin{matrix} {\cal N}_1 \\0\end{matrix}\right)\
,\ \ \ \ {\cal P}^2 = \left(\begin{matrix} 0\\
{\cal N}_2\end{matrix}\right)\ ,
\end{equation}
so that the two flavor components of the Higgs field  becomes
\begin{equation}
{ \phi}^1 = {\cal N}_1 \left(\begin{matrix} \chi_{11} M_{11}\\0
\end{matrix}\right) ,
\ \ \ \ \ \ \ \ {\phi^2} ={\cal N}_2 \left(\begin{matrix}
0\\M_{22}\end{matrix}\right)\ .
\end{equation}
with $M_{11}$ and $M_{22}$ the diagonal components of matrix $M$ to be still
determined.

This, together with equation (\ref{efes}), allows us to rewrite
equations (\ref{bogeqs1})-(\ref{bogeqs11}) in terms of strictly
periodic fields
\begin{eqnarray}
\tilde{F}^0_{ij} &=& - \left({\cal N}_1^2 |\chi_{11}|^2 M_{11}^2 + {\cal N}^2_2 M_{22}^2 - 2 + \frac{2\pi}{L_1L_2}\right)\varepsilon_{ij}\nonumber\\
\tilde{F}^3_{ij} &=&  \left(\kappa^2{\cal N}_1^2 |\chi_{11}|^2 M_{11}^2 -
\kappa^2{\cal N}^2_2 M_{22}^2 + \frac{2\pi}{L_1L_2}
\right)\varepsilon_{ij}\ . \label{modeleqs}
\end{eqnarray}

Using (\ref{efes}), (\ref{Atilde}) and (\ref{eme}), the periodic
piece of the field strength can be written as
\begin{eqnarray}
\tilde{F}_{ij}^0 &=& \varepsilon_{ij}\nabla^2 \eta^0\ \\
\tilde{F}_{ij}^3 &=&  \varepsilon_{ij}\nabla^2 \eta^1\
,\label{fieldstr}
\end{eqnarray}
where \be \nabla^2 = \partial_1^2 + \partial_2^2\ . \ee
Then, Bogomolnyi equations (\ref{modeleqs}) become
\begin{eqnarray}
\nabla^2  \eta^0 &=&  - {\cal N}_1^2 |\chi_{11}|^2 M_{11}^2 -
 {\cal N}^2_2 M_{22}^2 + 2 - \frac{2\pi}{L_1L_2}\label{lio1}\\
\nabla^2  \eta^1 &=&  \kappa^2{\cal N}_1^2 |\chi_{11}|^2 M_{11}^2 -
\kappa^2{\cal N}^2_2 M_{22}^2 + \frac{2\pi}{L_1L_2}\ .\label{lio2}
\end{eqnarray}
or using equation (\ref{eme})
\begin{eqnarray}
\nabla^2  \eta^0 &=& - {\cal N}_1^2 |\chi_{11}|^2 e^{- (\eta_0 - \eta_1)} -
 {\cal N}^2_2 e^{-(\eta_0 + \eta_1)} + 2 - \frac{2\pi}{L_1L_2}\label{ecuaciones2}\\
\nabla^2  \eta^1 &=&  \kappa^2 {\cal N}_1^2 |\chi_{11}|^2 e^{- (\eta_0 -
\eta_1)} - \kappa^2 {\cal N}^2_2 e^{-(\eta_0 + \eta_1)} +
\frac{2\pi}{L_1L_2}\ .\label{ecuaciones3}
\end{eqnarray}

From eqs. (\ref{ecuaciones2})-(\ref{ecuaciones3}) we see that \be
\nabla^2 \left(\frac{\eta_1}{\kappa^2} - \eta_0 \right)  =
  2 {\cal N}_1^2 |\chi_{11}|^2 e^{- (\eta_0 - \eta_1)}
 + \frac{2\pi}{L_1L_2} \left(\frac{1}{\kappa^2} +1\right) - 2
\ee
If we now integrate both sides on torus, the l.h.s. vanishes
since both $\eta_0$ and $\eta_1$ are periodic. We then find
\be
\int_{{\cal T}^2} d{\cal T} \left(
2 {\cal N}_1^2 |\chi_{11}|^2 e^{- (\eta_0 - \eta_1)}
 + \frac{2\pi}{L_1L_2} \left(\frac{1}{\kappa^2} +1\right) - 2\right) = 0
 \ee
Now,  since ${\cal N}_1^2 |\chi_{11}|^2 \exp(\eta_1 - \eta_0) \geq \
0$ one has

\begin{equation}
\left(\frac{\pi}{L_1L_2} \left(\frac{1}{\kappa^2} +1\right) - 1
\right) \int_{{\cal T}^2} d{\cal T}  \leq 0
\end{equation}
or, calling $A = L_1L_2$ the area of the torus, \be A \geq \pi
\left(1 + \frac{1}{\kappa^2}\right) \equiv A_c \ee That is, in order
to have consistent solutions from our ansatz, there
 is a minimal critical area which we call $A_c$, such that
 no solutions exists for  $A < A_c$.  It will be convenient to
introduce the parameter $\epsilon$,
\begin{equation}
\epsilon = 1 - \frac{A_c}{A}
\end{equation}
which measures the departure from this critical area, with range  $0 \leq
\epsilon \leq 1$.

We shall now solve the system
(\ref{ecuaciones2})-(\ref{ecuaciones3}) and, for simplicity, we
shall consider the case in which gauge coupling constants $e$ and
$g$ coincides so that $\kappa^2 = 1$ and the critical area reduces
to $A_c = 2\pi$  (the general case can be solved analogously).
Defining \be h^\pm = \frac{1}{2}(\eta_0 \pm \eta_1) \ee system
(\ref{ecuaciones2})-(\ref{ecuaciones3}) becomes
\begin{eqnarray}
\nabla^2  h^+ &=&  1 - {\cal N}_2^2\ e^{-2h^+} \label{deco12}\\
\nabla^2  h^- &=&  \epsilon - {\cal N}_1^2 |\chi_{11}|^2 e^{- 2
h^-} \ .\label{deco22}
\end{eqnarray}
We shall construct vortex configurations starting from the trivial
solution of eq.(\ref{deco12})
\begin{equation}
{\cal N}_2^2 = 1\ ,\ \ \ \ \ \ \ h^+ = 0\ .\label{que1}
\end{equation}

 Concerning eq.(\ref{deco22}), when
$\epsilon = 0$, it also has a trivial solution
\begin{equation}
{\cal N}_1^2 = 0\ ,\ \ \ \ \ \ \ h^- = 0.
\label{que2}
\end{equation}
which, together with (\ref{que1}) leads to $\eta_0=\eta_1 = 0$. For
$\epsilon \ne 0$ eq.(\ref{deco22}) can be solved extending the
method proposed in   \cite{GaR} for abelian vortices (see also
\cite{GaR2}, \cite{LMS}) which consists
 in expanding $h^-$ and ${\cal N}_1^2$ in powers of
 the Bradlow parameter $\varepsilon$,
 and further expand the fields in fourier
modes. The coefficients of the expansions obey recursive relations
that allow the numerical calculation of the magnetic and Higgs fields. This is
explained in an Appendix.

Using this method, we have solved numerically  eq.(\ref{deco22}) and obtained
${\cal N}_1$ and $h^-$ for different values of
$\epsilon$ in the range $0\leq \epsilon\leq 1$. From these results,
the magnetic field  $\tilde F^0_{12}$ and the Higgs fields $\Phi$ can be computed, using
  the equations
\begin{eqnarray}
\tilde F ^0_{12}(x_1,x_2) &=& \nabla^2 h^-(x_1,x_2) \label{magnet}
\end{eqnarray}
\begin{equation}
\Phi(x_1,x_2) =\left(
\begin{matrix}
{\cal N}_1\exp\left(i\pi \frac{(x_1 + i x_2) x_2}{{L_1 L_2} }
\right)\ \Theta(x_1,x_2)\ e^{-2{h^-}} & 0 \\
0 & 1
\end{matrix}
\right)\ ,\label{Phi}
\end{equation}
where $\Theta(x_1,x_2)$  is defined in (\ref{thetaa}).

\begin{figure}
\centering
\includegraphics[width=16cm]{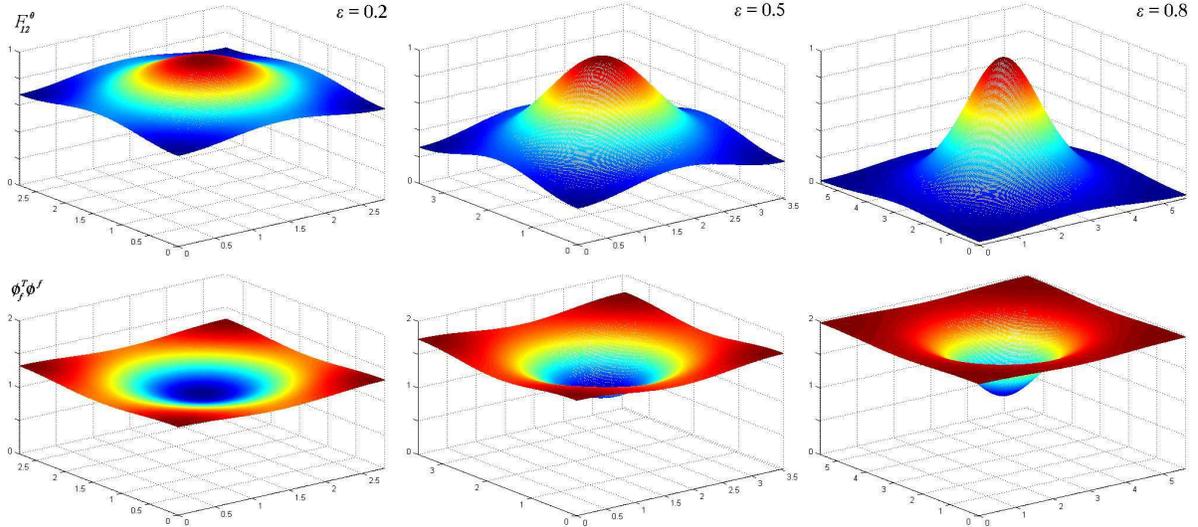}
\caption{\small  We plot $F^0_{12}$ and $\phi^\dag_f\phi^f$ for
different Areas. When $A = A_c$ ($\epsilon = 0$) the solutions are
trivial. When $A \to \infty$ ($\epsilon \to 1$) non-Abelian
vortices on the plane are recovered. We plot solutions for
different values of $\epsilon$. The area is written in units of
$\frac{2}{e^2\xi}$.} \label{Soluciones}
\end{figure}

We show some of
these solutions in Figure~\ref{Soluciones}. In all these cases,
$n_0=1$ so the flux $\int_{{\cal T}^2}d{\cal T} F^0_{12} = 2 \pi$ .
When varying the area $A_c \leq A \leq \infty$, solutions
interpolate continuously from the trivial constant solution for
$\epsilon = 0$ to the non-Abelian vortices on the plane for
$\epsilon = 1$. The vortex profiles are similar to those in the
plane, with the magnetic field concentrated around the position of
the vortex. At the center of the vortex, the upper component
($\phi^1$) of the Higgs field, the one with a non-trivial winding,
is zero,  as it happens in the Abelian case. Typically, when the
area is small $A\sim A_c$ the solutions converge fast, obtaining
high precision by computing a few orders of the $\epsilon$
expansion. In  the infinite area limit $A \gg A_c$ the method
converges    much slower. In this case we have considered up to 40
orders of the expansion with more than 400 Fourier modes (which
allows for a precision of less uncertainty than $10^{-6}$ for the
energy or magnetic flux).

\begin{figure}
\centering
\includegraphics[width=16cm]{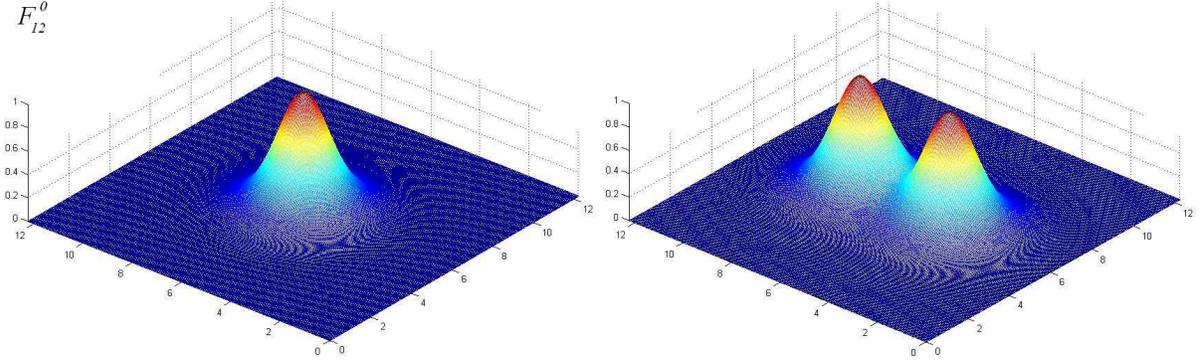}
\caption{\small  We plot elementary and non-elementary 2-vortex
configurations in the large area limit ($A \gg A_c$). The model has
gauge group $U(1)\times SU(2)$.} \label{plano}
\end{figure}

In the left panel of  Figure~\ref{plano} we show the magnetic
field for the $\epsilon \to 1$ case, which corresponds to an
elementary ($(1,0)$) non-Abelian vortex on the plane. The ansatz and the
numerical method work as well for the study of multi vortex
configurations, even when the vortices are not superimposed. We
show in the right panel of Figure~\ref{plano} a $(2,0)$-vortex
configuration in the limit $\epsilon \to 1$.

\section{$U(1)\times SU(N)$ strings}
In this section we extend the analysis  to the $N>2$ case, with  $
N_f = N$. We start from the Bogomolnyi equations
(\ref{Bogequa1})-(\ref{Bogequa2}) and consider a non-elementary
$(n,0,\dots,0)$ vortex
\begin{eqnarray}
F_{ij}^0 &=& - \left(\phi^\dag_f \phi^f -
N\right)\epsilon_{ij}\nonumber\\
F^m_{ij} &=& i \sqrt{2N} \kappa^2\phi^\dag_f
t^m\phi^f\epsilon_{ij}\ , \label{bogeqs2}
\end{eqnarray}
where again $\kappa$ is the ratio of the coupling constants, $\kappa
= g/e$. Other  elementary vortices like $(0,n,\dots,0)$, etc. can be
analogously treated.

We write the $N_f = N$ flavor multiplets $\phi^f_i$ in the form
\begin{eqnarray}
\phi^1_1 &=& {\cal N}_1 \exp\left(
-\frac{1}{\sqrt{2N}}\left(\eta^0 -
\sqrt{N-1}\eta^1\right)\right)\chi_{11} \nonumber\\
\phi^2_2 = \dots = \phi^N_N &=& {\cal N}_2
\exp\left(-\frac{1}{\sqrt{2N}}\left(\eta^0 +\frac{1}{\sqrt{N-1}}\
\eta^1 \right)\right)\ , \label{Efemonios}
\end{eqnarray}
where $\chi_{11}$ is defined in (\ref{chidef}), and $\eta^0$ and
$\eta^1$ (all other $\eta$'s are taken to be zero) are defined in
(\ref{eme}). With this ansatz, and defining
\begin{equation}
h^+ = \frac{1}{N}\left(\eta^0 + \frac{1}{\kappa^2\sqrt{N-1}}\
\eta^1\right)\ ,\ \ \ \ \ \ \ h^- = \frac{1}{N}\left(\eta^0 -
\frac{\sqrt{N-1}}{\kappa^2}\eta^1\right)
\end{equation}
we are able to rewrite equations (\ref{bogeqs2}) as
\begin{eqnarray}
\nabla^2 h^+ &=& \left(1 - \frac{\gamma^0 (1 -
\kappa^2)}{\kappa^2L_1L_2 N}\right) - {\cal N}_2^2
\exp\left(-\sqrt{\frac{2}{N}} \left((N - (1 - \kappa^2)) h^+ +(1 -
\kappa^2) h^-\right)\right)\label{prev1}\\
\nabla^2 h^- &=& \left(1 + \frac{\gamma^0}{L_1L_2 N} +
\frac{(N-1)\gamma^0}{\kappa^2L_1L_2N}\right)\nonumber\\
&-& {\cal N}_1^2 |\chi_{11}|^2 \exp\left(-\sqrt{\frac{2}{N}}
\left((N - 1)(1 - \kappa^2) h^+ + (1 + \kappa^2(N-1))
h^-\right)\right)\ ,\label{prev2}
\end{eqnarray}
where $\gamma^0$ is defined in (\ref{quince}).

As in the $N = 2$ case, we notice in equation (\ref{prev2}), that
since
\begin{equation}
\int_{{\cal T}^2} d{\cal T} \nabla^2 h^- = 0\ ,
\end{equation}
integrating equation (\ref{prev2}) on the torus gives
\begin{equation}
\left(1 + \frac{\gamma^0}{L_1L_2 N} +
\frac{(N-1)\gamma^0}{\kappa^2L_1L_2N}\right) \int_{{\cal T}^2}
d{\cal T} \geq 0 \ ,
\end{equation}
and this again implies the existence of a critical area $A_c$ such
that
\begin{equation}
A \geq A_c \equiv \frac{2\pi n}{N}\sqrt{\frac{2}{N}} \left(1 +
\frac{N-1}{\kappa^2}\right) \ .
\end{equation}

We again consider the parameter $ \epsilon = 1 -  {A_c}/{A}$ in
terms of which  equations (\ref{prev1})-(\ref{prev2}) read
\begin{eqnarray}
\nabla^2 h^+ &=& \left(1 + \frac{(1 - \epsilon)(1 - \kappa^2)}{N -
(1 - \kappa^2 )}\right) - {\cal N}_2^2 \exp\left(-\sqrt{2/N}
\left((N - (1 - \kappa^2)) h^+ +(1 -
\kappa^2) h^-\right)\right)\label{fin1}\\
\nabla^2 h^- &=& \epsilon - {\cal N}_1^2 |\chi_{11}|^2
\exp\left(-\sqrt{2/N} \left((N - 1)(1 - \kappa^2) h^+ + (1 +
\kappa^2(N-1)) h^-\right)\right)\ .\label{fin2}
\end{eqnarray}

For $\epsilon = 0$ one has the simple solutions
\begin{equation}
{\cal N}_1^2 = 0\ , \ \ \ \ {\cal N}_2^2 = \frac{N}{N - (1 - \kappa^2)} \label{eneord12}\\
\end{equation}
\begin{equation}
h^+ = h^- = 0\ .\label{etaM2}
\end{equation}
Non-trivial solutions when $\varepsilon \neq 0$ can be obtained as
before, Fourier expanding fields, and further expanding fields in
powers of $\varepsilon$. Order to order in $\epsilon$, one is left
with recursive relations for the coefficients. These relations can
be handled numerically  as in the $N=2$ case.

\section{Summary and discussion}

The main goal of this work was the study of  field configurations
corresponding to a periodic array of non-Abelian vortices. We have
considered a Yang-Mills theory coupled to fundamental scalar matter,
a model which can be seen as the truncated bosonic sector of a
${\cal N} = 2$ supersymmetric QCD.  We have studied these
configurations by solving the Bogomolnyi-Prasad-Sommerfeld equations
of the theory. By analyzing the (twisted) non-trivial boundary
conditions that the fields must satisfy on the two-torus, we were
able to propose an ansatz that reduces the BPS equations to a a
simpler  set of ordinary non linear equations that can be solved
numerically. These equations are solved perturbatively in powers of
a parameter measuring the departure of the area of the torus  from a
critical minimal value.

We have presented explicit solutions for the simplest gauge group
$U(2)$ which are the natural generalization of the ones studied by
Gonzalez Arroyo and Ramos \cite{GaR} to a non-Abelian Gauge
theory. On the other hand, for large areas, our solutions converge
to those studied in \cite{Auzzi}-\cite{GSY}, for SUSY QCD.

Our work could  be extended in several directions. We have analyzed
the case in which $N_f=N$. A natural extension would be to consider
the case in which $N_f
> N$ to study    non-Abelian semi-local
strings  \cite{ShifmanSemi} and this could be of interest in
connection with low-energy effective actions for string theories. It
is also natural to expect that the same ansatz presented here would
work practically in the same way for Chern-Simons-Matter theories
\cite{Aldrovandi:2007nb}-\cite{CSNAvor}, giving in this case origin
to configurations of periodic, electrically charged non-Abelian
vortices. Also, a similar analysis as the one presented here should
be of use to study non-Abelian periodic vortex array configurations
presenting BPS equations in the a non-Abelian model with adjoint
matter \cite{de Vega:1986eu}, \cite{CLS} or in the Standard Model
\cite{Bimonte:1994ik}. The  case considered here corresponds to the
particular set of parameters dictated by supersymmetry and BPS
equations. It is related, in the Abelian Higgs model,   to the limit
between Type I and Type II superconductivity, where vortices are
non-interacting. Away from this point, the full second order Euler
Lagrange equations should be solved. This case, that would
correspond  to interacting vortices, is technically more involved to
study. We expect that there exists a region in parameter space where
the vortex-vortex interaction is repulsive giving rise to a lattice
of vortices with a definite geometry. We hope to deal with some of
this issues in the future.

\vspace{1 cm}

\noindent\underline{Acknowledgements}: This work was partially
supported by ANPCYT (PICT 20204),  CICBA, CONICET (PIP 6160),
UBA and  UNLP (X310).
\newpage

\section*{Appendix}
To solve equation (\ref{deco22})
\begin{equation}
\nabla^2  h^- =  \epsilon - {\cal N}_1^2 |\chi_{11}|^2 e^{- 2
h^-}\ ,\label{ecuacion}
\end{equation}
we first define
\begin{equation}
H(x_1, x_2) = e^{2 h^-(x_1, x_2)}\ ,
\end{equation}
and rewrite (\ref{ecuacion}) in complex coordinates $z = x_1 + i
x_2$
\begin{equation}
\partial_z\left(H^{-1}\partial_{\bar z} H \right) =
\frac{1}{2}\left(\epsilon - {\cal N}_1^2 |\chi_{11}|^2
H^{-1}\right)\ .
\end{equation}

Since for $\varepsilon = 0$ there is a trivial solution \be H =
Constant,  \,\,\,\,\, {\cal N}_1 = 0\ , \ee and considering that
$0 \leq \epsilon \leq 1$, we can use $\varepsilon$ as a
perturbative parameter and expand $H$ and the normalization
constant ${\cal N}_1$ in powers of $\varepsilon$

\begin{equation}
H=\sum_{k=0}^{\infty}H_{k}\varepsilon^{k},  \ \ \ \
H^{-1}=\sum_{k=0}^{\infty}\bar{H}_{k}\varepsilon^{k}, \ \ \ \
{\cal N}_1^2=\sum_{k=0}^{\infty }N_{k}\varepsilon ^{k}\ .
\label{N2eps}
\end{equation}%
The coefficients $H_k$ and $\bar{H}_k$ are periodic functions and
can then be Fourier expanded
\begin{equation}
H_{k}=\ \sum_{n_{1}n_{2}} h_{n_{1}n_{2}}^{(k)}e^{2\pi i (n_1 x_1 /
L_1 + n_2 x_2 / L_2)}, \,\,\,\,\, \bar{H}_{k}=\
\sum_{n_{1}n_{2}}\bar{h}_{n_{1}n_{2}}^{(k)}e^{2\pi i (n_1 x_1 /
L_1 + n_2 x_2 / L_2)}\ ,
 \label{Hka}
\end{equation}
and the same can be done for $|\chi_{11}|^2$
\begin{equation}
|\chi_{11}|^2 =\ \sum_{n_{1}n_{2}} \eta_{n_{1}n_{2}} e^{2\pi i
(n_1 x_1 / L_1 + n_2 x_2 / L_2)} \ ,
\end{equation}
with normalized coefficients such that $\eta_{00} = 1$.

 Inserting these expansions in eq.(\ref{ecuacion}) one can determine
 order by order the coefficients,
\begin{eqnarray}
h_{n_{1}n_{2}}^{(0)} &=& \bar{h}_{n_{1}n_{2}}^{(0)}=  \left\{
\begin{array}{cc}
1 & n_{1}=\ n_{2}=0 \\ 0 & n_{1}\neq 0,n_{2}\neq 0
\end{array}
\right.
   \label{CoefHCero}
\nonumber\\
\nonumber\\
h_{n_{1}n_{2}}^{(1)}&=& \left\{
\begin{array}{cc}
0 & n_{1}=\ n_{2}=0 \\
\frac{2\pi \eta_{n_{1}n_{2}}}{|\xi _{n_{1}n_{2}}|^2} & n_{1}\neq 0,n_{2}\neq 0%
\end{array}%
\right.   \label{CoefHUno}
\nonumber\\
\nonumber\\
\bar{h}_{n_{1}n_{2}}^{(1)} &=& - h_{n_{1}n_{2}}^{(1)}\ \ \ \ \ \ \
\ \ \ \ \ \ \ \ \ \ \ \ \ \ \ \ \ \ \ \ \ \ \ \ \ \ \ \ \ \ \ \ ,
\end{eqnarray}
where
\begin{equation}
\xi_{n_{1}n_{2}}\equiv \pi \sqrt{T} \left(i n_{1} +
\frac{n_{2}}{\sqrt{T}} \right)\ , \label{epsDeriv}
\end{equation}
with ${T} = L_{2}/L_{1}$ the aspect ratio of the torus. In the
same way one can calculate coefficients to any order $Q$ in
$\varepsilon$
\begin{equation}
h_{n_{1}n_{2}}^{(Q)}=\left \{
\begin{array}{cc}
0 & n_{1}=\ n_{2}=0 \\
\frac{C_{n_{1}n_{2}}^{(A)}-C_{n_{1}n_{2}}^{(B)}-C_{n_{1}n_{2}}^{(C)}}{
|\xi_{n_{1}n_{2}} |^2} & n_{1}\neq 0,n_{2}\neq 0%
\end{array}%
\right.   \ ,\label{CoefHN}
\end{equation}
with
\begin{eqnarray}
C_{q_{1}q_{2}}^{(A)}\!\!\!&=& \!\!\!\sum_{n_{1}n_{2}}
\sum_{k=1}^{Q-1}\ \bar{h}_{n_{1}n_{2}}^{(k)}\
h_{q_{1}-n_{1},q_{2}-n_{2}}^{(Q-k)}\ \xi _{q_{1}q_{2}}\ \bar{\xi
}_{q_{1}-n_{1},q_{2}-n_{2}} \nonumber
\\
C_{q_{1}q_{2}}^{(B)}\!\!\!&=&\!\!\! 2\pi \sum_{n_{1}n_{2}}
\sum_{k=0}^{Q-1}\ \bar{h}_{n_{1}n_{2}}^{(k)}\ N_{Q-k}\ \eta
_{q_{1}-n_{1},q_{2}-n_{2}} \nonumber
\\
C_{q_{1}q_{2}}^{(C)}\!\!\!&=& \!\!\!\sum_{n_{1}n_{2}}
\sum_{k=0}^{Q-2}\ \bar{h}_{n_{1}n_{2}}^{(k)}\
h_{q_{1}-n_{1},q_{2}-n_{2}}^{(Q-k-1)}\ \xi _{q_{1}q_{2}}\ \bar{\xi
}_{q_{1}-n_{1},q_{2}-n_{2}} \ .\nonumber\label{CB}
\end{eqnarray}
Coefficients ${\bar h}_{n_{1}n_{2}}$, appearing in the expansion
of $H^{-1}$, are obtained from the condition $HH^{-1} = 1$
\begin{equation}
\bar{h}^{(Q)}_{q_{1} q_{2}} = - \sum_{n_{1}n_{2}} \sum_{k =
1}^{Q}\  h^{(k)}_{n_{1}n_{2}}\ \bar{h}^{(Q-k)}_{q_{1} - n_{1},
q_{2} - n_{2}}\ .
\end{equation}

One also has  to find a recurrence relation for the coefficients
$N_k$. For this, the condition $\int_{{\cal T}^2} d{\cal T}
\tilde{F}_{12} = 0$ implies
\begin{eqnarray}
N_0 &=& 0 \nonumber\ , \ \ \ \ \ \ \
N_1 = 1 \nonumber\\
N_Q &=& - \sum_{n_{1}n_{2}} \sum_{k=1}^{Q-1}\
\eta_{-n_{1},-n_{2}}\ \bar{h}_{n1n2}^{(k)}\ N_{Q-k}  \ , \;\;\; Q>1 \ .\nonumber\\
\label{AA}
\end{eqnarray}

Computing these recursive relations we can obtain $h^-$ and ${\cal
N}_1$, and with this compute the magnetic and Higgs field from
eqs.(\ref{magnet})-(\ref{Phi}).


\newpage
\begin{thebibliography}{99}
\bibitem{N}
  H.~B.~Nielsen and P.~Olesen,
  Nucl.\ Phys.\  B {\bf 61}, 45 (1973).
\bibitem{bogo}
  E.~B.~Bogomolnyi,
  Sov.\ J.\ Nucl.\ Phys.\  {\bf 24} (1976) 449
  [Yad.\ Fiz.\  {\bf 24} (1976) 861].
\bibitem{deVega}
  H.~J.~de Vega and F.~A.~Schaposnik,
  Phys.\ Rev.\  D {\bf 14} (1976) 1100.
\bibitem{taubes} C.~H.~Taubes, Comm.\ Math.\ Phys.\ {\bf 72}
(1980) 277.
\bibitem{bradlow} S. B. Bradlow,
Commun. Math. Phys. 135 (1990) 1–17.
\bibitem{GaR}
  A.~Gonzalez-Arroyo and A.~Ramos,
  JHEP {\bf 0407} (2004) 008
\bibitem{lecTong}
  D.~Tong,
Lectures  at  TASI 2005,
  arXiv:hep-th/0509216.
\bibitem{fas} F.A.~Schaposnik,
Lectures at 4th Advanced Chilean School of Astrophysics, Cosmology and
Gravitation, Valparaiso, Dec 2006.
  arXiv:hep-th/0611028.

\bibitem{dVS2}H.~J.~de Vega and F.~A.~Schaposnik,
  Phys.\ Rev.\ Lett.\  {\bf 56} (1986) 2564;
  Phys.\ Rev.\  D {\bf 34} (1986) 3206.
\bibitem{Hanany1}
  A.~Hanany and D.~Tong,
  JHEP {\bf 0307}, 037 (2003)
  [arXiv:hep-th/0306150].
  \bibitem{Auzzi}
  R.~Auzzi, S.~Bolognesi, J.~Evslin, K.~Konishi and A.~Yung,
  Nucl.\ Phys.\  B {\bf 673} (2003) 187
\bibitem{Hanany2}
  A.~Hanany and D.~Tong,
  JHEP {\bf 0404}, 066 (2004)
  \bibitem{SY} M.~Shifman and A.~Yung,
   Phys.\ Rev.\  D {\bf 70} (2004) 045004
\bibitem{GSY}
  A.~Gorsky, M.~Shifman and A.~Yung,
  Phys.\ Rev.\  D {\bf 71} (2005) 045010
\bibitem{mod1}
  M.~Eto, Y.~Isozumi, M.~Nitta, K.~Ohashi and N.~Sakai,
  Phys.\ Rev.\ Lett.\  {\bf 96} (2006) 161601
\bibitem{mod2}
  M.~Eto, Y.~Isozumi, M.~Nitta, K.~Ohashi and N.~Sakai,
  J.\ Phys.\ A  {\bf 39} (2006) R315
\bibitem{revshif} M.~Shifman and A.~Yung,
arXiv:hep-th/0703267.
\bibitem{GA2}
  A.~Gonzalez-Arroyo, Talk  at  the Summer School on Nonperturbative Quantum Field Physics,
  Peniscola, Spain,
  arXiv:hep-th/9807108.
\bibitem{tHooft}
  G.~'t Hooft,
  Nucl.\ Phys.\  B {\bf 153} (1979) 141.
   \bibitem{ENS}
  J.~D.~Edelstein, C.~Nunez and F.~Schaposnik,
  Phys.\ Lett.\  B {\bf 329} (1994) 39
  \bibitem{GaR2} A.~Gonzalez-Arroyo and A.~Ramos,
   JHEP {\bf 0701} (2007) 054
  \bibitem{LMS}P. Forgacs, G.S. Lozano, E.F. Moreno, F.A. Schaposnik,
JHEP {\bf 0507} (2005) 074;
   G.~S.~Lozano, D.~Marques and F.~A.~Schaposnik,
  JHEP {\bf 0609}, 044 (2006)
\bibitem{ShifmanSemi}
  M.~Shifman and A.~Yung,
  Phys.\ Rev.\  D {\bf 73} (2006) 125012
\bibitem{Aldrovandi:2007nb}
  L.~G.~Aldrovandi and F.~A.~Schaposnik,
  arXiv:hep-th/0702209.
\bibitem{CSNAvor}
  G.~S.~Lozano, D.~Marques, E.~F.~Moreno and F.~A.~Schaposnik,
  arXiv:0704.2224 [hep-th].
\bibitem{de Vega:1986eu}
  H.~J.~de Vega and F.~A.~Schaposnik,
  Phys.\ Rev.\ Lett.\  {\bf 56} (1986) 2564.
  \bibitem{Bimonte:1994ik}J.~Ambjorn and P.~Olesen, Int.\ J.\ Mod.\ Phys.\ {\bf
  A5} (1990) 4525; Nucl.\ Phys.\ {\bf B330}, (1990) 193;
  G.~Bimonte and G.~Lozano,
  Phys.\ Rev.\  D {\bf 50} (1994) 6046; Y. Yang,
  Physica {\bf D101} (1997) 57.
  \bibitem{CLS}L.F.Cugliandolo, G.S.Lozano, F.A.Schaposnik, Phys.\ Rev.\ {\bf D40}(1989) 3440.
\end{thebibliography}
\end{document}